\AtBeginDocument{%
  \paperwidth=\dimexpr
    1in + \oddsidemargin
    + \textwidth
    + 1in + \oddsidemargin
  \relax
  \paperheight=\dimexpr
    1in + \topmargin
    + \headheight + \headsep
    + \textheight
    + 0.7in + \topmargin
  \topmargin = -0.5in
  \relax
  \usepackage[pass]{geometry}\relax
}

\RequirePackage{fix-cm}
\documentclass{svjour3}                     
\smartqed  
\usepackage{graphicx}
\usepackage{graphicx}
\usepackage{dcolumn}
\usepackage{bm}
\usepackage{dsfont}
\usepackage{amsmath}
\usepackage{amssymb}
\usepackage[usenames]{color}
\usepackage[colorlinks=true,linkcolor=blue,citecolor=blue,urlcolor=blue]{hyperref}
\usepackage{cite}

\def\mS{\ensuremath{\mathcal{S}}}
\def\mD{\ensuremath{\mathcal{D}}}

\def\mC{\ensuremath{\mathcal{C}}}
\def\mF{\ensuremath{\mathcal{F}}}

\newcommand{\gS}[1]{#1\!\!\!\!\!\not~}

\newcommand{\pslash}{\gS{p}}

\begin{document}

\title{Four-quark states from functional methods
}


\author{Gernot Eichmann \and Christian S. Fischer \and Walter Heupel \and Nico Santowsky \and Paul C. Wallbott}



\institute{  G. Eichmann \at
				LIP Lisboa, Av.~Prof.~Gama~Pinto 2,  1649-003 Lisboa, Portugal\\
                Departamento de F\'isica, Instituto Superior T\'ecnico, 1049-001 Lisboa, Portugal \\
				\email{gernot.eichmann@tecnico.ulisboa.pt}
          \and
				C. S. Fischer, W. Heupel, N. Santowsky, P. C. Wallbott \at
              Institut f\"ur Theoretische Physik \\
              Justus-Liebig Universit\"at Gie{\ss}en\\
              35392 Gie{\ss}en, Germany
            \and
           C. S. Fischer \at
           Helmholtz Forschungsakademie Hessen f\"ur FAIR (HFHF),\\
            GSI Helmholtzzentrum f\"ur Schwerionenforschung, Campus Gie{\ss}en,\\
             35392 Gie{\ss}en, Germany\\
              \email{Christian.Fischer@theo.physik.uni-giessen.de}
}

\date{Received: date / Accepted: date}

\maketitle

\begin{abstract}
In this feature article we summarise and highlight aspects of the treatment of four-quark states
with functional methods. Model approaches to those exotic mesons almost inevitably have to assume certain
internal structures, e.g. by grouping quarks and antiquarks into (anti-)diquark clusters or
heavy-light $q\bar{q}$ pairs. Functional methods using Dyson-Schwinger and Bethe-Salpeter
equations can be formulated without such prejudice and therefore have the potential to put these assumptions
to test and discriminate between such models. So far, functional methods have been used to study the
light scalar-meson sector and the heavy-light sector with a pair of charmed and a pair of light quarks in
different quantum number channels. For all these states, the dominant components in terms of internal two-body clustering
have been identified. It turns out that chiral symmetry breaking plays an important role for the
dominant clusters in the light meson sector (in particular for the scalar mesons) and that this property
is carried over to the heavy-light sector. 
Diquark-antidiquark components,
on the other hand, turn out to be almost negligible for most states with the exception of open-charm
heavy-light exotics.
\keywords{exotic mesons \and tetraquarks \and Bethe-Salpeter equations}
\end{abstract}

\section{Introduction}
\label{sec:intro}

One of the spectacular successes of this millennium's hadron spectroscopy experiments is the discovery
of many 'exotic' meson states in the heavy-quark energy region that do not seem to fit into the conventional
$q\bar{q}$ picture, see e.g.
\cite{Chen:2016qju,Lebed:2016hpi,Esposito:2016noz,Guo:2017jvc,Ali:2017jda,Olsen:2017bmm,Liu:2019zoy,Brambilla:2019esw}
for recent review articles. Some of these carry net electromagnetic charge and thus may be naturally
explained as four-quark states $Q\bar{Q}q\bar{q}$ (with $q=u,d,s$ and $Q=c,b$). As a consequence,
four-quark states are generally considered as promising candidates to explain the structure and
properties of these exotic hadrons.

On the other hand, four-quark states may also play an important role in the low-energy spectrum of QCD.
The lightest multiplet of scalar mesons has been under debate for a long time. Its experimental situation appeared
questionable and its place in the Review of Particle Physics \cite{Zyla:PDG2020} has been firmly established  only in this century. Since then,
uncertainties in the mass and width of the $\sigma/f_0(500)$ have been steadily reduced \cite{Caprini:2005zr,Yndurain:2007qm,GarciaMartin:2011jx,Moussallam:2011zg}.
The notion that the light scalar meson nonet is incompatible with a conventional $q\bar{q}$ picture, however,
goes back some way \cite{Jaffe:1976ig}: by assuming a dominant four-quark structure, characteristic properties
like inverted mass hierarchies and decay patterns are naturally explained.

One of the most interesting questions concerning four-quark states is the nature of their internal structure formed
by the dynamics of QCD. There is extensive literature available on this subject, however a general consensus
is not in sight. Effective theories and model approaches study the spectrum that emerges from specific
configurations, which may be grouped into three categories:
   \begin{itemize}\setlength\itemsep{1mm}
   \item The quarks and antiquarks may cluster into internal
   \textit{diquark-antidiquark} 
   pairs that interact via coloured forces, see
   e.g.~\cite{Esposito:2016noz} for a review.\footnote{We will use the terms \textit{four-quark} and \textit{tetraquark} synonymously,
   although in the literature the latter often refers to diquark-antidiquark configurations only.}

   \item The \textit{hadro-quarkonium} picture~\cite{Voloshin:2007dx},
   mostly relevant in the charm and bottom regions, suggests a heavy quark and antiquark grouped together in a
   tight core and surrounded by a light $q\bar{q}$ pair.

   \item The \textit{meson-molecule} picture of arrangements
   into pairs of heavy-light mesons is especially relevant for states close to meson-meson thresholds,
   see e.g.~\cite{Guo:2017jvc} for a review.
   \end{itemize}

It is important to note that these possibilities are not mutually exclusive: In general, every experimental
state may be a superposition of components with a different structure and the `leading' component may be different
on a case-by-case basis. It is therefore of utmost importance to develop theoretical approaches to QCD that can
deal with and distinguish between all these possibilities, such as the lattice and the functional approach to QCD.
Lattice QCD has made important and interesting progress so far, see
\cite{Prelovsek:2010kg,Prelovsek:2013cra,Abdel-Rehim:2014zwa,Lee:2014uta,Prelovsek:2014swa,Padmanath:2015era,Francis:2016hui,Bicudo:2017szl,Francis:2018jyb,Leskovec:2019ioa} and references therein.
In this feature article we focus on functional methods and summarise results from recent works on this topic
\cite{Heupel:2012ua,Eichmann:2015cra,Wallbott:2019dng,Wallbott:2020jzh}.

The article is organized as follows. In the next section we collect a number of general remarks on the functional
framework of Dyson-Schwinger and Bethe-Salpeter equations (DSEs and BSEs) and outline some of the properties of the four-quark
equation that will be important in what follows. We also discuss the properties of the BSE for $q\bar{q}$ states
with respect to chiral symmetry breaking and briefly summarize the pseudo-Goldstone boson nature of the pion.
In Sec.~\ref{sec:3} we focus on the light scalar meson octet. We explain how two different formulations
of the four-body BSE can be used to shed light on the internal structure of the $f_0(500)$ and its partners and
highlight the important role chiral symmetry breaking plays for those states. In Sec.~\ref{sec:4} we discuss recent results on
heavy-light four-quark states, in particular those with hidden or open-charm content.
We conclude with a summary and outlook in Sec.~\ref{sec:5}.

\section{Bethe-Salpeter equations}
\label{sec:BSE}

In this section we highlight some aspects of functional methods, in particular DSEs and BSEs,
that are important in connection with four-quark states. A pedagogical introduction to the subject can be found in
the review article \cite{Eichmann:2016yit} and many practical and technical details are collected in \cite{Sanchis-Alepuz:2017jjd}.

\begin{figure*}[t]
	\begin{center}
		\includegraphics[width=\textwidth]{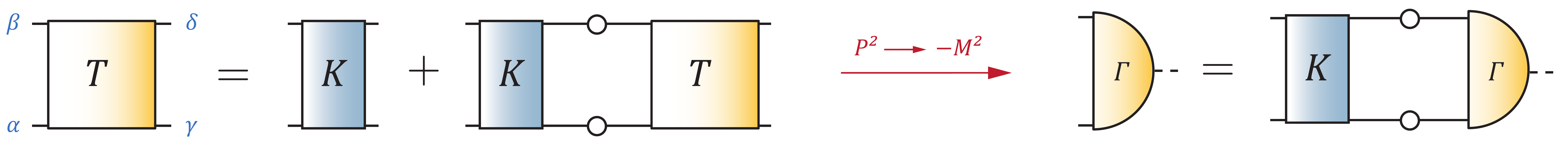}
		\caption{Dyson equation~\eqref{dyson-eq} and onshell Bethe-Salpeter equation~\eqref{bse-general} in graphical form. }\label{fig:bse}
	\end{center}
\end{figure*}

\subsection{The BSE for quark-antiquark states}\label{sec:2-body-bse}

We begin with the homogenuous BSE for states that can be built from a $q\bar{q}$ pair.
As illustrated in Fig.~\ref{fig:bse}, it is extracted from the (exact)
Dyson equation for the quark four-point function $\mathbf{G}_{\alpha\beta,\delta\gamma}$.
We drop indices and momentum integrations in the notation\footnote{We will frequently use such a compact notation
but emphasise that it is only for notational convenience.} and denote the disconnected product of a dressed quark and
(anti-)quark propagator by $\mathbf{G}_0$. The amputated and connected part of $\mathbf{G}$ is the
scattering matrix $\mathbf{T}$ defined by $\mathbf{G} = \mathbf{G}_0 + \mathbf{G}_0\, \mathbf{T}\,\mathbf{G}_0$.
Both satisfy Dyson equations:
\begin{align}\label{dyson-eq}
\mathbf{G} = \mathbf{G}_0 + \mathbf{G}_0\,\mathbf{K}\,  \mathbf{G}  \qquad  \Leftrightarrow \qquad
\mathbf{T} = \mathbf{K}   + \mathbf{K}\,  \mathbf{G}_0\,\mathbf{T}\;.
\end{align}
It is important to note that at this stage these four-point functions are exact and therefore contain all possible meson poles,
including virtual bound states or resonances with pole locations on the second Riemann sheet with respect to the total momentum $P^2$
of the $q\bar{q}$ pair.

At the respective pole locations, $\mathbf{G}$ and $\mathbf{T}$ can be written as
\begin{align}\label{pole-in-G-or-T}
\mathbf{G} \rightarrow \frac{\mathbf{\Psi}\,\overline{\mathbf{\Psi}}}{P^2+M^2} \qquad \Leftrightarrow \qquad \mathbf{T} \rightarrow \frac{\mathbf\Gamma\,\overline{\mathbf\Gamma}}{P^2+M^2}\,,
\end{align}
where the Bethe-Salpeter amplitude (BSA) $\mathbf\Gamma$ appears together with its conjugate as the residue of the onshell meson pole at $P^2 = -M^2$.
The corresponding BS wave function $\mathbf{\Psi} = \mathbf{G}_0\,\mathbf\Gamma$
can be viewed as the analogue of a quantum-mechanical wave function in relativistic quantum field theory
and contains the full momentum, Dirac-Lorentz, color and flavor structure of the state.
Comparing residues on both sides of the equation yields the homogeneous BSE at the pole (see Fig.~\ref{fig:bse}):
\begin{align}\label{bse-general}
\mathbf\Gamma = \mathbf{K}\,\mathbf{G}_0\,\mathbf\Gamma \,.
\end{align}
The homogeneous BSE always comes together with a normalization condition, which also automatically enforces electromagnetic charge normalization
without further conditions, see \cite{Eichmann:2016yit} for details.

In what follows, one of the most important properties of the BSE for $q\bar{q}$ states is its behaviour in the chiral limit where all
current-quark masses vanish. Here, the pseudoscalar mesons play a special role since they are the Goldstone bosons in QCD's chiral symmetry
breaking pattern. It can be rigorously shown that they emerge as such from their BSEs, see \cite{Eichmann:2016yit,Horn:2016rip} for details
and references. In practical calculations, this property needs to be preserved by the approximations and truncations one invokes.

To this end we need to discuss the
DSE for the quark propagator, which provides one of the major ingredients in every calculation involving BSEs. Written
in the same shorthand notation, the DSE reads
\begin{align}\label{quark-dse}
S^{-1} = S_0^{-1} + g^2 \Gamma^{qqg}_0 S D \Gamma^{qqg}\,,
\end{align}
where $S$ denotes the dressed quark propagator, $D$ the dressed gluon, $\Gamma^{qqg}$ the dressed quark-gluon vertex and corresponding
bare quantities are denoted by subscript zero. The running coupling of QCD is denoted by $g^2$ and various factors and momentum dependencies
are kept implicit. This DSE captures the effect of dynamical mass generation due to dynamical chiral symmetry
breaking, i.e. the resulting masses $M(p^2)$ of the dressed light quarks $S(p) = Z_f(p^2) \left(i \pslash + M(p^2)\right)^{-1} $ are
in the range of $M(p^2=0) \sim 400$ MeV and therefore two orders of magnitude larger than the current-quark masses.

The self-energy of the
quark DSE and the interaction kernel $\mathbf{K}$ of the meson BSE are related by the axialvector Ward-Takahashi identity (axWTI), which
is dictated by chiral symmetry. Any approximation of the quark DSE and meson BSE that maintains this identity reproduces the
\mbox{(pseudo-)} Goldstone boson nature of the lightest pseudoscalar-meson octet. Thus in the chiral limit, the axWTI-controlled kernel $\mathbf{K}$
provides exactly the amount of binding energy needed to obtain a massless state from two massive constituents.

In practice, a well-explored and frequently used truncation satisfying this property is the rainbow-ladder approach, which amounts
to the replacement
\begin{align}\label{RL}
g^2 D  \Gamma^{qqg} \rightarrow \frac{\alpha(k^2)}{k^2} \Gamma^{qqg}_0\,,
\end{align}
where $k^2$ is the squared gluon momentum and $\alpha(k^2)$ plays the role of an effective running coupling. All results discussed herein have
been obtained using a particular version of this truncation, the Maris-Tandy model \cite{Maris:1999nt}. Beyond-rainbow-ladder solutions
of the quark DSE and corresponding results for the meson spectrum are available, see e.g. \cite{Williams:2015cvx}.
Overviews and more references can be found in \cite{Bashir:2012fs,Cloet:2013jya,Eichmann:2016yit}.

The rainbow-ladder truncation of the meson BSE is particularly adequate for ground-state pseudoscalar and vector mesons.
By extension, scalar and axialvector diquarks satisfy analogous BSEs and provide the underlying dynamics
in ground-state octet and decuplet baryons~\cite{Eichmann:2016yit}. This will be important in what follows since the two-body clusters
in the four-quark states under consideration are those of pseudoscalar and vector mesons as well as scalar and axialvector diquarks.

\begin{table}
\caption{Rainbow-ladder meson and diquark masses (in MeV; $n=u,d$).
		$m_q$ is the input current-quark mass at a renormalization point $\mu=19$ GeV in a momentum-subtraction scheme.
		The column $m_{PS}$ contains  the masses of $\pi$, $D$, $D_s$, $\eta_c$ and
		the column $m_V$ those of $\rho/\omega$, $D^*$, $D_s^*$ and $J/\psi$.
		The columns $m_S$ and $m_A$ list the respective diquark masses. The quoted errors
		are obtained by varying the parameter in the rainbow-ladder interaction. }
\label{tab:rl}
\vspace{-2mm}
\begin{center}
\begin{tabular}{cr@{\qquad\qquad}ccc@{\qquad\qquad}ccc}
\hline\noalign{\smallskip}
                        & $m_q$     &                   &     $m_{PS}$    & $m_V$     &              &$m_S$    & $m_A$ \\
\noalign{\smallskip}\hline\noalign{\smallskip}
		            $n$ &    3.7    & $n\bar{n}$        &          \hphantom{1} 138\hphantom{(0)}   & \hphantom{0}732(1)\hphantom{0}   &  $nn$        &\hphantom{0}802(77) &  \hphantom{0}999(60) \\
		           $s$ &    91      & $c\bar{n}$        &        1802(2)  & 2068(16)  &  $nc$        &2532(90) & 2572(8)\hphantom{0}  \\
		            $c$ &    795    & $c\bar{s}$        &        1911(3)  & 2169(14)  &  $sc$        &2627(82) & 2666(7)\hphantom{0}  \\
		             &              & $c\bar{c}$        &        2792(6)  & 2980(6)\hphantom{0}   &  $cc$        &3382(15) & 3423(8)\hphantom{0}   \\
\noalign{\smallskip}\hline
\end{tabular}
\end{center}
\vspace{-2mm}
\end{table}

 For later reference, the rainbow-ladder meson and diquark masses are collected in Table~\ref{tab:rl}.
 Working in the isospin symmetric limit, the $u/d$ current-quark mass is fixed by $m_\pi$,
 the charm-quark mass by the sum $m_D+m_{D^*}$ and the strange-quark mass by $m_{D_s}+m_{D_s^*}$.
 The deviations between theory and experiment are then below $7\%$ for all meson masses,
 although the mass splittings in rainbow-ladder are quite large
 (see~\cite{Maris:2005tt,Rojas:2014aka,Gomez-Rocha:2014vsa,Fischer:2014xha,Fischer:2014cfa} for further discussions).
 Note that the diquarks are considerably heavier than the respective mesons ---
 a scalar diquark made of light $u/d$ quarks comes out at about 800 MeV and an axialvector diquark at 1 GeV.
 This property extends to the heavy-quark region, where the diquarks are roughly 500 -- 700 MeV heavier than their meson partners.

\begin{figure}[t!]
	\centerline{%
		\includegraphics[width=\textwidth]{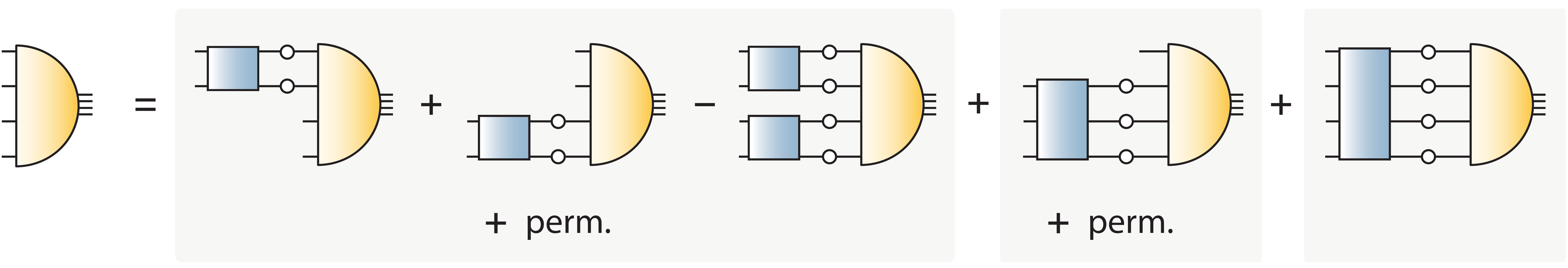}}
	\caption{Four-body BSE for the tetraquark Bethe-Salpeter amplitude.}
	\label{fig:4b-bse}
\end{figure}

\subsection{The BSE for four-quark states}\label{sec:2.2}

The BSE for a bound state or resonance built from two quarks and two antiquarks ('four-quark BSE') is derived analogously
from the quark eight-point function $\mathbf{G}_{\alpha\beta\lambda\sigma,\delta\gamma\rho\tau}$.
The details of the corresponding interaction kernel $\mathbf{K}$ have been worked out in \cite{Huang:1974cd,Khvedelidze:1991qb,Yokojima:1993np}.
As shown in Fig.~\ref{fig:4b-bse}, it contains (from right to left) irreducible four-body interactions,
irreducible three-body interactions, and three terms containing irreducible two-body interactions. Here, the third term
on the right hand side of the equation is
necessary to avoid over-counting due to the iteration of the first two terms.
The two-body kernels are the same as those in the $q\bar{q}$ system shown in Fig.~\ref{fig:bse}.

So far the irreducible three- and four-body interactions have not been included in practical calculations.
On the one hand, this is due to practical feasibility, but there are arguments for the dominance of two-body correlations
through the appearance of poles associated with internal two-body clusters. We will discuss this point in more detail at the end of the present subsection.

\begin{figure}[t!]
	\centerline{ %
		\includegraphics[width=0.8\textwidth]{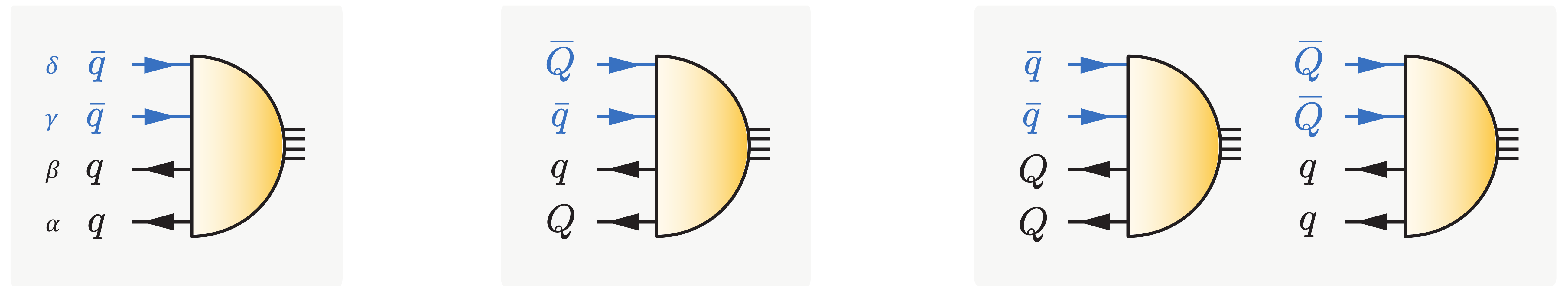}}
	\caption{Different configurations for four-quark states: four equal quarks ($qq\overline{qq}$), hidden flavor ($Qq\overline{q}\overline{Q}$)
             and open flavor ($QQ\overline{qq}$, $qq\overline{QQ}$). }
	\label{fig:3confs}
\end{figure}

   Apart from the interaction kernel, the essential ingredient of the four-quark BSE is the
   four-body BSA $\mathbf\Gamma_{\alpha\beta\gamma\delta}(p_1, p_2, p_3, p_4)$,
   which contains the full momentum, Dirac-Lorentz, color and flavor structure of the state.
   Denoting light and heavy quark flavors by $q$ $(u,d,s)$ and $Q$ ($c$, $b$), one can distinguish
   the systems shown in Fig.~\ref{fig:3confs}:
   states made of four equally massive quarks ($qq\overline{qq}$);
   those with a heavy $Q\bar{Q}$ and light $q\bar{q}$ pair (hidden flavor); and those with a heavy $QQ$ and light $\overline{qq}$ pair
   or vice versa (open flavor).
   Counting the quarks from bottom to top,
   (12)(34) then corresponds to diquark-antidiquark-like configurations and (13)(24) and (14)(23) to meson-meson-like configurations.
   The greek subscripts absorb all Dirac, color and flavor indices and
   the total momentum of the four-quark state is $P = p_1 + p_2 + p_3 + p_4$.

   Let us briefly sketch the construction of the BSA (details can be found in Ref.~\cite{Wallbott:2020jzh}).
   To this end, one constructs all possible color, flavor and Dirac-momentum tensors and
   combines them such that they satisfy the relevant symmetries of the system.
   These are the Pauli antisymmetry in (12) or (34),
   \begin{align}
   \mathbf\Gamma(p_2, p_1, p_3, p_4)_{\beta\alpha\gamma\delta} \stackrel{!}{=}  &- \mathbf\Gamma(p_1, p_2, p_3, p_4)_{\alpha\beta\gamma\delta}\,, \label{pauli-12} \\
   \mathbf\Gamma(p_1, p_2, p_4, p_3)_{\alpha\beta\delta\gamma} \stackrel{!}{=}  &- \mathbf\Gamma(p_1, p_2, p_3, p_4)_{\alpha\beta\gamma\delta}\,, \label{pauli-34}
   \end{align}
   where a permutation of all Dirac, color and flavor indices is understood,
   and charge-conjugation symmetry in (13)(24) or (14)(23):
   \begin{align}
   C_{\alpha\alpha'}\,C_{\beta\beta'}\,C_{\gamma\gamma'}\,C_{\delta\delta'}\,\mathbf\Gamma(-p_3,-p_4,-p_1,-p_2)_{\gamma'\delta'\alpha'\beta'}
    & \stackrel{!}{=} \pm \mathbf\Gamma(p_1, p_2, p_3, p_4)_{\alpha\beta\gamma\delta}\,, \label{cc-13-24} \\
   C_{\alpha\alpha'}\,C_{\beta\beta'}\,C_{\gamma\gamma'}\,C_{\delta\delta'} \,\mathbf\Gamma(-p_4,-p_3,-p_2,-p_1)_{\delta'\gamma'\beta'\alpha'}
    & \stackrel{!}{=} \pm \mathbf\Gamma(p_1, p_2, p_3, p_4)_{\alpha\beta\gamma\delta}\,, \label{cc-14-23}
   \end{align}
   where $C = \gamma_4 \gamma_2$ is the charge-conjugation matrix and the signs $\pm$ determine the $C$ parity of the state.
   The amplitude for a $qq\overline{qq}$ system satisfies all four relations, whereas
   a hidden-flavor amplitude $Qq\overline{q}\overline{Q}$ is only subject to Eq.~\eqref{cc-14-23} and
   an open-flavor configuration $QQ\overline{qq}$ only satisfies Eqs.~(\ref{pauli-12}--\ref{pauli-34}).

   The color part of the amplitude
   is made of two independent color-singlet tensors, which can be taken from
   the (13)(24) and (14)(23) configurations in the meson-meson  ($\mathbf{1} \otimes \mathbf{1}$) channels:
   \begin{equation}\label{color-11}
   (\mC_{11})_{ABCD} =  \frac{1}{3}\,\delta_{AC}\,\delta_{BD} \,, \qquad
   (\mC_{11}')_{ABCD} =  \frac{1}{3}\,\delta_{AD}\,\delta_{BC}\,.
   \end{equation}
   By Fierz identities, any other possible color tensor is a linear combination of these,
   like in the (12)(34) diquark-antidiquark decomposition ($\mathbf{\bar{3}} \otimes \mathbf{3}$):
   \begin{equation}\label{color-33}
   \mC_{\bar{3}3} = -\frac{\sqrt{3}}{2}\,(\mC_{11}-\mC_{11}')  \quad \text{with} \quad
   (\mC_{\bar{3}3})_{ABCD} = -\frac{1}{2\sqrt{3}}\, \varepsilon_{ABE}\,\varepsilon_{CDE}\,.
   \end{equation}

   The flavor part of the amplitude follows from the Clebsch-Gordan construction.
   For example, a system made of four light quarks $q \in \{u,d\}$ satisfies all symmetries~(\ref{pauli-12}--\ref{cc-14-23}).
   From $\mathbf{2}\otimes \mathbf{2} \otimes \mathbf{\bar{2}} \otimes \mathbf{\bar{2}}=$ $(\mathbf{1}_a \oplus \mathbf{3}_s) \otimes (\mathbf{1}_a \oplus \mathbf{3}_s)$
   one obtains 16 flavor wave functions with definite Pauli symmetry, which can be arranged into isospin multiplets:
   \begin{equation}
       \mF_{aa}\,(I=0), \qquad
       \mF_{ss}\,(I = 0, 1, 2)\,, \qquad
       \mF_{as}\,(I=1)\,, \qquad
       \mF_{sa}\,(I=1)\,.
   \end{equation}
    Here, $\mF_{aa} = [ud][\bar{u}\bar{d}]$ is the $I=0$ flavor wave function typically associated with the $\sigma$ meson which
    is antisymmetric  in both (12) and (34).

   With four Dirac indices for the quarks, the Dirac-momentum part of $\mathbf\Gamma$
   consists of all linearly independent tensors permitted by Lorentz invariance.
   For example, a scalar tetraquark with quantum numbers $J^{PC} = 0^{++}$ features the tensors
   \begin{equation} \label{tensors} \renewcommand{\arraystretch}{1.4}
      \begin{array}{c}
        \gamma^5_{\alpha\gamma}\,\gamma^5_{\beta\delta}\,, \\
        \gamma^5_{\alpha\delta}\,\gamma^5_{\beta\gamma}\,,
      \end{array} \qquad
      \begin{array}{c}
        \gamma^\mu_{\alpha\gamma}\,\gamma^\mu_{\beta\delta}\,, \\
        \gamma^\mu_{\alpha\delta}\,\gamma^\mu_{\beta\gamma}\,,
      \end{array}\qquad
      \begin{array}{c}
        (\gamma_5 C)_{\alpha\beta}\,(C^T \gamma_5)_{\gamma\delta}\,, \\
        (\gamma^\mu C)_{\alpha\beta}\,(C^T \gamma^\mu)_{\gamma\delta}\,,
      \end{array}\qquad
      \begin{array}{c}
        \delta_{\alpha\gamma}\,\delta_{\beta\delta}\,, \\
        \delta_{\alpha\delta}\,\delta_{\beta\gamma}\,,
      \end{array} \qquad
             \dots
   \end{equation}
   where Greek subscripts are Dirac indices only.
   Combined with all momenta in the system and taking into account Fierz identities,
   there are 256 such linearly independent tensors for (pseudo-)scalar states~\cite{Eichmann:2015cra}, whereas
   for (axial-)vector states there are $3 \times 256 = 768$~\cite{Wallbott:2019dng}.
   The (momentum-dependent) strengths of these tensors are the dynamical outcome
   of the four-body BSE and therefore provide information on the overlap with the state in question.
   The dominant tensors are the `$s$-wave' tensors that do not depend on any relative momentum
   and thus also do not carry orbital angular momentum.
   For $J^{PC} = 0^{++}$ there are 16 $s$-wave tensors which form a Fierz-complete subset.

   Combined with appropriate color and flavor wave functions,
   one can see that the tensors in Eq.~\eqref{tensors} have overlap with systems made of two pseudoscalar mesons,
   two vector mesons, two scalar diquarks, or two scalar mesons, respectively. For example, the combination
   \begin{equation}
        \left( \gamma^5_{\alpha\gamma}\,\gamma^5_{\beta\delta}\,\mC_{11} + \gamma^5_{\alpha\delta}\,\gamma^5_{\beta\gamma}\,\mC_{11}' \right) \mF_{aa}
   \end{equation}
   satisfies all symmetries~(\ref{pauli-12}--\ref{cc-14-23}) and contributes to the BSA of the $\sigma$ meson.
   Moreover, it also coincides with the quantum numbers of two pseudoscalar mesons inside the four-quark state.
   In this way one can construct a physically motivated subset
   of tensors that correspond to the possible two-body clusters and systematically proceed until  the tensor basis is complete.

   Finally, the aforementioned momentum-dependent strengths are the
   Lorentz-invariant dressing functions of the BSA,
   which depend on all Lorentz invariants in the system. From four independent momenta one can construct 10 Lorentz invariants,
   one of which ($P^2=-M^2$) is fixed on the mass shell. The remaining 9 invariants are dynamical, which illustrates the numerical complication
   in solving the four-quark BSE --- an analogous meson BSE features only two dynamical variables and a baryon's three-quark BSE five~\cite{Eichmann:2016yit}.
   It is advantageous to group these 9 variables into multiplets of the permutation-group $S_4$,
   which yields a singlet $\mS_0$, a doublet $\mD$ and two triplets~\cite{Eichmann:2015nra,Eichmann:2015cra}. The singlet and doublet are given by
   \begin{equation}\label{momentum-multiplets}
      4\mS_0 =  \sum_{i=1}^4 p_i^2 - \frac{P^2}{4}\,,  \quad
      \mD = \frac{1}{4\mS_0}\left[ \begin{array}{c} \sqrt{3}\,(p_1-p_2)\cdot (p_3-p_4) \\ (p_1+p_2)\cdot (p_3+p_4) - 2 (p_1\cdot p_2 + p_3\cdot p_4) \end{array}\right].
   \end{equation}

\begin{figure}[t!]
	\centerline{%
		\includegraphics[width=1\textwidth]{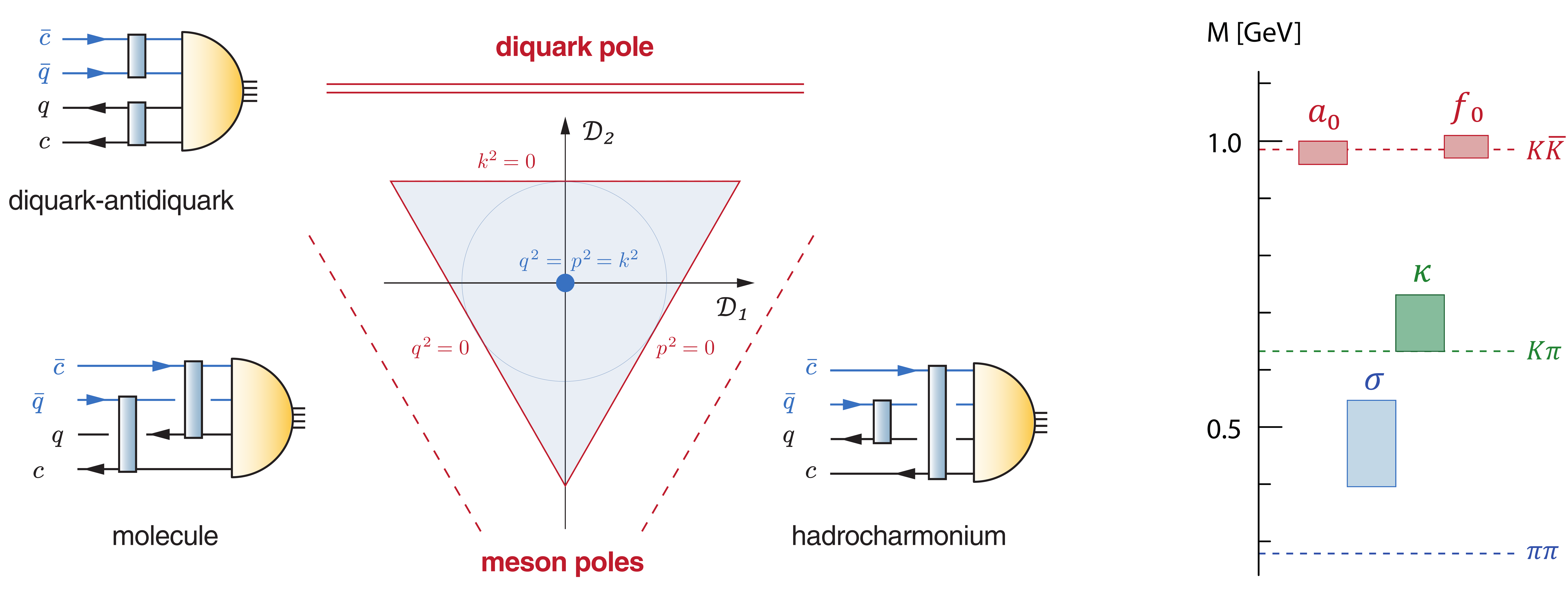}}
	\caption{Left: Singularities within the four-body BSE. Right: Mass ordering in the light scalar nonet from the PDG~\cite{Zyla:PDG2020}.}
	\label{fig:poles}
\end{figure}

   The doublet has an important property that is illustrated in the left of Fig.~\ref{fig:poles}.
   Its two variables form a plane from where one can identify intermediate two-body  (meson-meson, diquark-antidiquark) poles
   emerging in the solution of the equation.
In the plot the locations of the singularities are denoted
by dashed lines (for $q\bar{q}$ poles, i.e. meson clusters) and a double line (for poles in diquark clusters). The integration
region of the BSE is indicated by the shaded triangle. Depending on the putative masses of the internal clusters, these lines may lie far outside
the integration region, close by or, in case of resonances, even inside the integration region. Clearly, in the first case they hardly affect
the BSE, whereas in the two latter cases they leave noticeable traces. Depending on the quantum numbers of the four-quark state and those
of the putative internal clusters, it is a dynamical question which of the clusters is the most important. This is
the reason why the four-quark BSE can be used to discriminate between the three different possibilities of internal clustering discussed in the
introduction and indicated in Fig.~\ref{fig:poles}. We will frequently come back to this point throughout the article.

If strong two-body correlations inside the four-quark state exist that lead to clustering with poles close to or inside the integration region,
then these will dominate the BSE. In particular, it is plausible that these effects also dominate the analytic background provided by the terms
with irreducible three- or four-body interactions. This argument, and the numerical complexity potentially induced by these terms, explains why
the latter have been neglected so far \cite{Heupel:2012ua,Eichmann:2015cra,Wallbott:2019dng,Wallbott:2020jzh}.

  We conclude on the general note that
  whatever the internal structure of a particular state is, any gauge-invariant bound-state or resonance pole
  with matching meson quantum numbers must show up in the quark eight-point function $\mathbf{G}_{\alpha\beta\lambda\sigma,\delta\gamma\rho\tau}$,
  whose residue is the four-quark BSA determined by its BSE.
  Because the same can also be said about the quark four-point function $\mathbf{G}_{\alpha\beta,\delta\gamma}$ and the $q\bar{q}$ BSE derived from it,
  it is conceivable that \textit{without} approximations both equations should produce the same spectrum.
  Any possible internal structure (molecule, hadrocharmonium, diquark-antidiquark, $q\bar{q}$, hybrid, $\dots$)
  would then be encoded in the respective interaction kernel and/or the BSA itself.
  Under approximations it becomes a question of practicability what the best starting point is:
  if a state is dominated by $qq\overline{qq}$, an approximate four-quark BSE
  will produce a more reliable result than an approximate $q\bar{q}$ BSE
  (where four-quark effects would need to enter through a complicated interaction kernel).
  Therefore, what the four-quark BSE studies discussed herein can accomplish at the present stage
  is to disentangle the relative importance of different $qq\overline{qq}$ components in a given state,
  in particular the two-body clusters formed by molecular-like, hadroquarkonium and diquark-antidiquark configurations.

\section{Light exotic four-quark states}
\label{sec:3}

\subsection{The physics case}
\label{sec:3.1}

In the following we briefly outline the physics case of the light scalar meson nonet with quantum numbers $J^{PC} = 0^{++}$; a detailed and executive discussion
can be found  in the review article \cite{Pelaez:2015qba}.

The experimental evidence for the existence of a light scalar meson nonet below
1 GeV is compelling and has led to more and more precise values for the masses and widths in current issues of the
Review of Particle Physics \cite{Zyla:PDG2020}. Similarly uncontroversial seems the general notion that the members
of this nonet cannot be thought of as ordinary mesons, i.e. states  primarily consisting of $q\bar{q}$ constituents.
The main points can already be inferred from the experimental spectrum (right panel of Fig.~\ref{fig:poles}).
If the light scalar mesons were ordinary $q\bar{q}$ states, the isosinglet $\sigma/f_0(500)$ should be mass-degenerate
with the isotriplet $a_0(980)$, followed by the kaon-like $\kappa/K_0^\ast(700)$ and the $f_0(980)$ as the $s\bar{s}$ state;
but what happens instead is that the $a_0(980)$ and $f_0(980)$ are (almost) mass-degenerate.
Second, in the non-relativistic quark model the scalar mesons are $p$ waves and carry orbital angular momentum,
so they should lie above $1$ GeV like their axialvector and tensor meson partners with quantum numbers $1^{+-}$, $1^{++}$ and $2^{++}$.
Third, whereas the $a_0(980)$ and $f_0(980)$ sit close to the $K\bar{K}$ threshold and are narrow, the $\sigma$ and $\kappa$ are broad states.
This ties in with the fact that the experimentally observed
$\sigma$ cannot be described by a simple Breit-Wigner ansatz, see \cite{Giacosa:2007bn} and Refs. therein.

Instead, there seems general agreement that the dominant structure of the $f_0(500)$ and its multiplet partners is that of
four-quark ($qq\overline{qq}$) states.
This picture offers a natural explanation for the inverted mass hierarchy of the light scalar nonet
(the $a_0(980)$ and $f_0(980)$ now have the same quark content) and its decay pattern with
large decay widths for the $\sigma$ and $\kappa$, which can fall apart into $\pi\pi$ and $K\pi$~\cite{Jaffe:1976ig,Jaffe:2004ph}.
In addition, strong support for the dominant $qq\overline{qq}$ nature
of these states comes from considerations of the $1/N_c$ expansion, both around $N_c=3$ and in the large-$N_c$ limit
\cite{Pelaez:2003dy,Pelaez:2006nj,Nebreda:2011cp,RuizdeElvira:2010cs}. In particular, it was shown that the pole location
of the $\rho$ meson follows a distinctive dependence on $N_c$ as expected for $q\bar{q}$ states, whereas the pole locations
of scalar mesons do not. This also ties in with the fact that the mass pattern of the light scalar mesons does not allow
for accommodation in the usual Regge trajectories characteristic for ordinary mesons \cite{Anisovich:2000kxa}.

The internal structure of the light scalar mesons is less clear. In principle, one could envisage three different possibilities
for the four (anti-)quarks to arrange. If irreducible four-body forces were dominant inside the four-quark state (e.g., mediated
by a four-gluon vertex), then
one would expect a tightly bound state without internal clustering. If irreducible two-body forces were dominant, one might
expect an internal clustering either in diquark-antidiquark configurations \cite{Jaffe:1976ig,Hooft:2008we} or
meson-meson pairs \cite{Weinstein:1990gu}. Which of these latter two possibilities
is the more important one depends on the dynamics of the two-body forces and needs to be investigated. This has been one of the
major goals in studies using functional methods \cite{Heupel:2012ua,Eichmann:2015cra}.

Certainly, it is by no means excluded that there is some $q\bar{q}$ admixture in the physical states of the light scalar nonet,
although judging from the experimental spectrum in Fig.~\ref{fig:poles} it is likely small.
In fact, evidence from several theoretical sources points towards this direction \cite{Black:1998wt,Close:2002zu,Giacosa:2006tf,Hooft:2008we,Londergan:2013dza,Pelaez:2006nj,Nieves:2011gb,RuizdeElvira:2010cs}.
Within functional methods this has been studied by some of us in a very recent work \cite{Santowsky:2020pwd}. Here
we will not touch upon this issue but refer the interested reader to the original articles.

In the next subsection we briefly outline the two different formulations of the
four-body equation that have been used so far with functional methods
and in Sec.~\ref{sec:3.3} we summarise results on the light scalar meson nonet.

\subsection{Light four-quark states in the four-body and two-body formalism}
\label{sec:3.2}

\begin{figure}[t!]
		\includegraphics[width=1\textwidth]{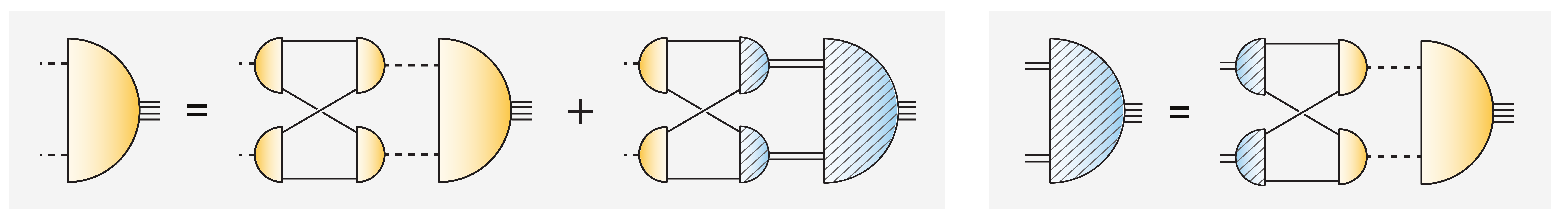}
	\caption{Two-body equation in terms of meson (yellow) and diquark (hatched, blue) degrees of freedom.
             Meson and diquark propagators are denoted by dashed and double lines, respectively.}
	\label{fig:4body-2body}
\end{figure}

At present, light four-quark states have been investigated within functional methods using two approaches:
\begin{itemize}\setlength\itemsep{1mm}

\item The first is the original four-body BSE shown in Fig.~\ref{fig:4b-bse}. As detailed above, it is solved
      by neglecting the influence of irreducible three- and four-body forces.
      This approximation is justified if two-body interactions are the driving force for internal clustering inside the four-quark states.
      In practice the equation is solved by reducing the tensor basis to the physically dominant $s$-wave tensors and restricting the momentum
      dependence to the singlet and doublet variables in Eq.~\eqref{momentum-multiplets}.
      In the following we refer to this as the \textit{four-body approach}~\cite{Eichmann:2015cra}.

\item Provided that two-body correlations are dominant, one may even go further and simplify the system to a two-body equation.
      The derivation proceeds through a Faddeev-Yakubowski equation~\cite{Yakubovsky:1966ue},
      where the two-quark scattering matrices are represented in terms of meson and diquark poles. The resulting equation
      is shown in Fig.~\ref{fig:4body-2body} and expressed in terms of meson-meson and diquark-antidiquark BSAs.
      In the following we call this the \textit{two-body approach} and refer to Ref.~\cite{Heupel:2012ua} for technical details.
\end{itemize}
   Note that the two-body approach is already quite close in spirit to a description in an effective field theory,
   although the interaction kernel is not expressed through effective meson exchanges.
   Instead, the resulting coupled system of equations describes meson-meson and diquark-antidiquark
   components that interact with each other by quark exchange,
   with the same dressed quark propagator determined from its DSE~\eqref{quark-dse}.
   One of the great advantages of this formulation is the simplicity of the BSAs since all Dirac structure is gone
   and one has to deal with much fewer momentum variables.
   In turn, one needs to determine the meson and diquark BSAs as well as their propagators that enter in the equation
   from their respective BSEs. These are solved using the rainbow-ladder approximation (\ref{RL}) for the quark-(anti)quark
   interaction described in Sec.~\ref{sec:2-body-bse} and thus no additional model input is needed. As we will see below,
   the results from the four-body and two-body approach are very similar.

   Finally, let us note that the simplification to a two-body system is analogous to the \textit{quark-diquark} approach for baryons.
   Also in that case one can study baryons either from the original three-quark BSE or from the corresponding quark-diquark BSE.
   In both cases the results are very similar, which suggests a dominance of two-quark correlations in terms of diquarks~\cite{Eichmann:2016yit,Eichmann:2016hgl,Chen:2017pse}.
   Nevertheless, the dynamics that drive the four-quark system are quite different compared to baryons
   since diquark correlations are now accompanied by meson-meson correlations.
   In fact, one can see that only one diagram appears on the r.h.s. of the second equation in Fig.~\ref{fig:4body-2body} because
   one cannot exchange quarks between a diquark and an antidiquark (unless one allows for fully disconnected diagrams in the derivation~\cite{Kvinikhidze:2014yqa}).
   Thus, one can merge both equations into a single one for the meson-meson amplitude.
   As a result, already the structure of the equation implies that the system
   is driven by meson-meson configurations, whereas diquark degrees of freedom appear only internally in higher loop diagrams.

\subsection{Results for the lightest scalar meson octet: $f_0(500)$ and company}
\label{sec:3.3}

\begin{figure}[t!]
	\includegraphics[width=1\textwidth]{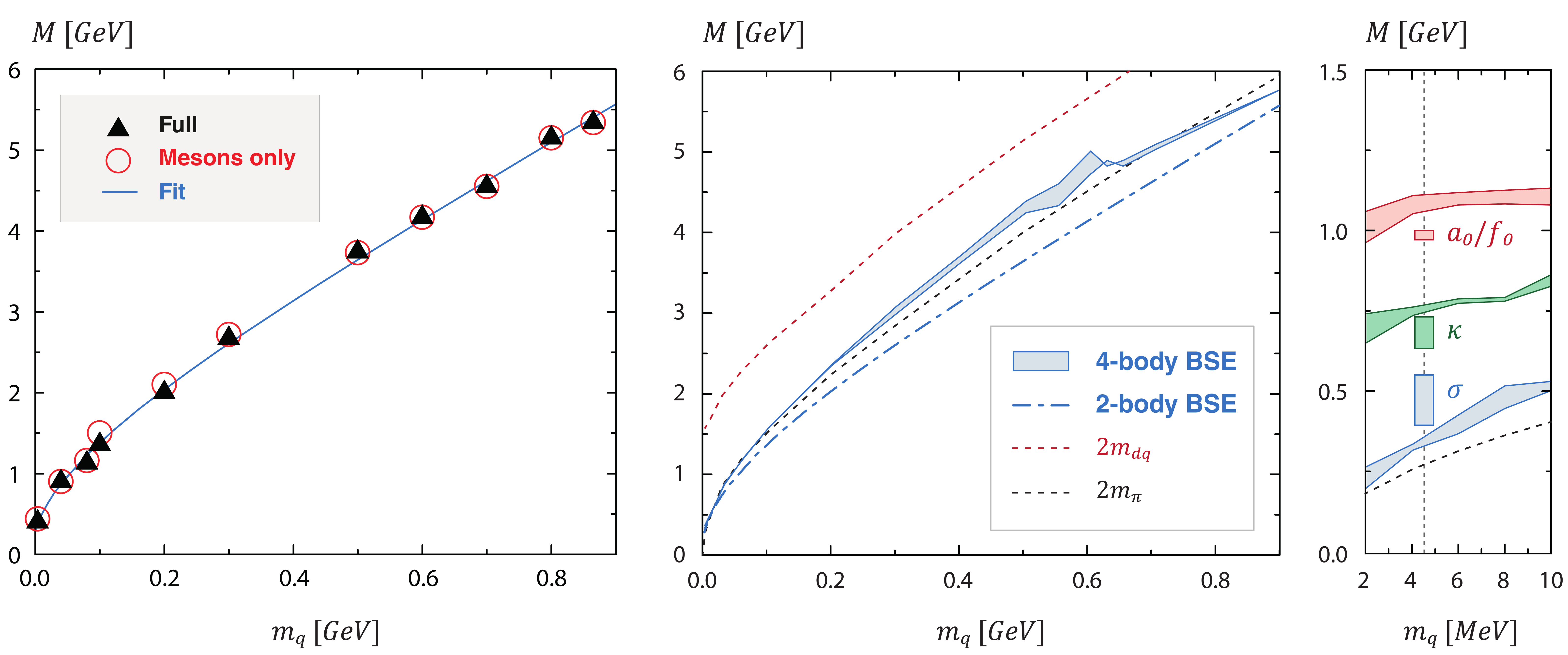}
	\caption{Left: Results for the light isosinglet scalar meson from the effective two-body approach \cite{Heupel:2012ua}.
		     Center and right: Results from the four-body approach \cite{Eichmann:2015cra}.}
	\label{fig:lightresults}
\end{figure}

Let us first discuss the results in the two-body approach~\cite{Heupel:2012ua} shown in the left panel of Fig.~\ref{fig:lightresults}.
For a scalar four-quark state with quantum numbers $0^{++}$,
the lightest possible and thus dominant components are pseudoscalar mesons ($0^{-+}$) and scalar diquarks ($0^+$).
The plot shows the mass of the light isosinglet scalar meson as a function of the current
quark mass $m_q$ (same for all four quarks).
Compared are the masses obtained from the full system with those taking into account only the $\pi\pi$ components.
There are two notable observations: First, the contributions from scalar diquarks are negligible and the scalar state is completely dominated by the $\pi\pi$
component. This is a direct consequence of dynamical chiral symmetry breaking, which makes the mass of the pion much lighter
than the typical mass scale of the scalar diquark (which is of the order of 800 MeV). Second, except for the offset in the chiral limit,
the mass evolution of the four-quark state is similar to that of the pion. Since the latter is governed by the Gell-Mann-Oakes-Renner relation,
 $m_{\text{PS}}^2 \sim m_q$, so is the mass of the four-quark state as can be seen in the plot in the center.
 Clearly, dynamical chiral symmetry breaking plays an important role for the mass evolution and the structure of the light scalar mesons.

A very similar result has been obtained in the four-body formalism~\cite{Eichmann:2015cra}. The center panel of Fig.~\ref{fig:lightresults}
shows that the mass evolutions from the four-body and two-body approach are qualitatively similar.
Also qualitatively similar, however quantitatively different is the relation to the $\pi\pi$
threshold defined by $2m_{\text{PS}}$. Whereas in the effective two-body approach the result stays below the $\pi\pi$ threshold in the heavy mass range and only crosses
the threshold shortly before the physical point, in the four-body approach this crossing happens at much larger masses around $m_q=700$ MeV.
This probably indicates that truncation effects in either formulation currently prohibit a precision determination of this scale.

The right panel of Fig.~\ref{fig:lightresults}
shows the results for the light scalar meson nonet for different current-quark masses compared with the experimental values at
the physical point, indicated by the vertical dashed line. The error bands reflect variations of the model parameter for the effective coupling,
Eq.~\eqref{RL}. Clearly, there is qualitative agreement between the experimental and the theoretical spectrum. (Part of) the quantitative
difference between theory and experiment might be related to the fact that we neglected part of the tensor structure of the four-body
BSA to make the equation tractable for a numerical setup. Another part might come from missing $q\bar{q}$ contributions, which have
not been included in either calculation. With respect to the second point, the very recent result of
\begin{equation}
M_{0^{++}} = 456(24) \,\mbox{MeV}
\end{equation}
for the real part of the mass of the $f_0(500)$ in an approach where the two-body equation in Fig.~\ref{fig:4body-2body}
has been coupled to the $q\bar{q}$ BSE is encouraging~\cite{Santowsky:2020pwd}.

In any case, the explicit calculation in the four-body framework illustrates how a dynamical generation of resonances emerges from the
quark level (see e.g. \cite{Oset:2016lyh} for a review of other approaches): Pseudoscalar-meson poles are generated in the process of
solving the four-body BSE and dominate the dynamics of the system.
If they enter the integration domain of the BSE (for current quark masses below about 700 MeV), the four-body state becomes a resonance.
It is amusing to note what happens if the possibility for the dynamical formation of internal clusters is cut out from the BSE. This
can be done by neglecting the dependence of the BSAs on the total momenta of the internal two-body states, i.e., the
kinematic variables arranged in the permutation-group doublet~\eqref{momentum-multiplets}.
With the very same two-body interaction one then obtains a bound state  with a mass of the order of
1500 MeV at the physical point. This is just what one would expect for a bound state of four massive quarks with an infrared (constituent-like) quark
mass function $M(p^2=0) \sim 400$ MeV. By contrast, if one allows for internal clustering into pseudo-Goldstone bosons by
using a BSA rich enough in structure to depend on the doublet variables, the mass goes down into the 400 MeV range
displayed in Fig.~\ref{fig:lightresults}. This mechanism, together with the towering dominance of $\pi\pi$ clusters over diquark-antidiquark
contributions, are the most important results of Refs.~\cite{Heupel:2012ua,Eichmann:2015cra}.

\begin{figure*}[t]
\includegraphics[width=1\textwidth]{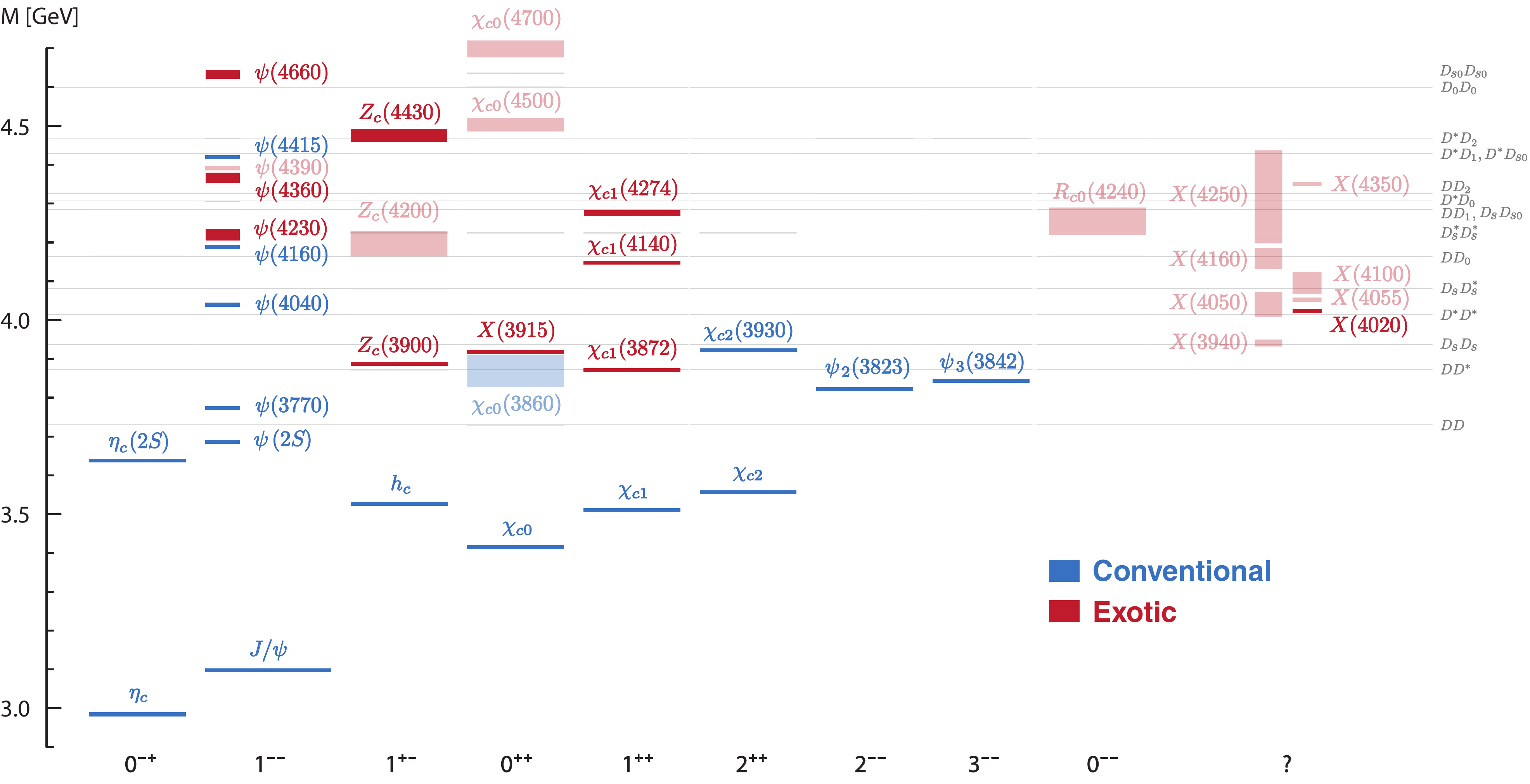}
\caption{Charmonium spectrum as of July 2020 from the PDG~\cite{Zyla:PDG2020}.
Not well-established states are shown in pale colors and the box heights are the mass ranges.
The open-charm thresholds are shown in gray (we drop the bars for $D\bar{D}$, $D\bar{D}^\ast$, $\dots$ for notational clarity).}  \label{fig:charmonium}
\end{figure*}

\section{Heavy-light exotic four-quark states}
\label{sec:4}

\subsection{The physics case}
\label{sec:4.2}

  While chiral symmetry and relativity complicate the properties of light-quark systems,
  the heavy-quark sector is a much cleaner environment for studying exotic mesons.
  Here already the success of the non-relativistic quark model provides
  a basic measure for distinguishing `ordinary' from `exotic' mesons, namely through
  the deviation from the quark-model expectations.
  Following a series of discoveries over the past two decades,
  a number of exotic meson candidates in the charmonium region (the `$XYZ$ states') are experimentally well-established by now
  (Fig.~\ref{fig:charmonium}):
  \begin{itemize} \setlength\itemsep{1mm}
  \item The $\chi_{c1}(3872)$ or $X(3872)$ was first reported by Belle in 2003~\cite{Choi:2003ue} and the
        first exotic charmonium-like state to be found.
        The fact that its mass is indistinguishable from the $D^0 \bar{D}^{\ast 0}$ threshold,
        and that its narrow width ($<1.2$ MeV) is about two orders of magnitude smaller than potential-model predictions for the excited
        $c\bar{c}$ state $\chi_{c1}'$, makes it difficult to reconcile with a conventional charmonium picture.
        The suppression of its $J/\psi\,\gamma$ decay mode compared to $J/\psi\,\pi^+\pi^-$
        and the isospin violation in the $J/\psi\,\rho$ decay
        are also unusual for a $c\bar{c}$ state. The $X(3872)$ has been seen in a number of reactions and
        its quantum numbers are $I(J^{PC}) = 0(1^{++})$.
        Other exotic candidates in this channel are the $X(4140)$ and $X(4274)$ which are seen in the $J/\pi\,\phi$ mass spectrum.

  \item The $\psi(4230)$, now identified with the $Y(4260)$, is one of several exotic candidates in the $1^{--}$ vector channel which are accessible in $e^+ e^-$ collisions.
        Their exotic assignment is mainly due to the overpopulation of the $1^{--}$ channel with regard to well-established  $c\bar{c}$ states
        and the fact that in contrast to ordinary charmonia they are less likely to decay into open-charm final states.

  \item The $Z_c(3900)$ and $Z_c(4430)$ with $1^{+-}$ carry charge and are thus manifestly exotic
        since their minimal quark content is $c\bar{c}u\bar{d}$.

  \end{itemize}

  Several other exotic states in the charmonium spectrum are well established,
  such as the $X(3915)$ with quantum numbers $0^{++}$ (although a $2^{++}$ identification is also possible).
  In the bottomonium sector there are currently two exotic candidates, the $Z_b(10610)$ and $Z_b(10650)$ with quantum numbers $1^{+-}$,
  and a possible $cc\overline{cc}$ candidate has recently been reported by LHCb~\cite{Aaij:2020fnh}.

  Even though some signals may be kinematical threshold effects,
  the question whether four-quark states exist \textit{at all} is no longer truly disputed in light of the current experimental evidence.
  The main objective is therefore to understand their internal structure.
  The theoretical interpretations range from hadronic molecules to hadrocharmonia, diquark-antidiquark states
  and hybrid mesons; see the reviews~\cite{Chen:2016qju,Lebed:2016hpi,Esposito:2016noz,Guo:2017jvc,Ali:2017jda,Olsen:2017bmm,Liu:2019zoy,Brambilla:2019esw} for detailed discussions and an overview of theoretical approaches.
  Of course, quantum field-theoretically all these configurations can mix together as well as with
  ordinary $q\bar{q}$ states, but it is conceivable that certain configurations are dominant for particular states, like for example
   the proximity to some threshold is a typical signal for a molecule.
  This is also the question that studies with functional methods have been attempting to answer based upon
  internal quark-gluon dynamics. In the following we will give a brief overview on the existing results.

\subsection{Results in the four-body formalism}
\label{sec:4.3}

   The existing work on heavy-light four-quark states using functional methods
   is based on the four-body approach of Fig.~\ref{fig:4b-bse},
   both for systems with a heavy $Q\bar{Q}$  (hidden flavor) or $QQ$ pair (open flavor)~\cite{Wallbott:2019dng,Wallbott:2020jzh}. 
   The construction of their BSAs follows the outline in Sec.~\ref{sec:2.2}
   and is detailed in Ref.~\cite{Wallbott:2020jzh}.

   Let us focus on the hidden-flavor case and in particular $cq\overline{qc}$ systems with hidden charm,
   since those comprise most of the experimentally established tetraquark candidates to this date.
   In this case,
   out of the possible symmetries~(\ref{pauli-12}--\ref{cc-14-23}) the amplitude  only satisfies charge conjugation in (13)(24), Eq.~\eqref{cc-14-23}.
   The three configurations (13)(24), (14)(23) and (12)(34) correspond to those
   displayed in Fig.~\ref{fig:poles}:
   \begin{itemize}\setlength\itemsep{1mm}
   \item (13)(24), $(c\bar{q})(q\bar{c})$, molecule: $(D,D^\ast,\dots)\times(D,D^\ast,\dots)$
   \item (14)(23), $(c\bar{c})(q\bar{q})$, hadrocharmonium: $( J/\psi, \eta_c, \dots)\times(\pi,\rho,\eta,\omega, \dots)$
   \item (12)(34), $(cq)(\overline{qc})$, diquark-antidiquark: $(S,A,\dots)\times(S,A,\dots)$
   \end{itemize}
   The possible two-body clusters
   are further narrowed down by the $J^{PC}$ and isospin quantum numbers of the state in question.

   As illustrated in Fig.~\ref{fig:poles}, the two-body clusters dynamically produce poles in the BSA.
   The pole locations correlate with the respective thresholds, i.e., the sums of the two-body masses.
   Therefore, the proximity of such a pole to the integration region, indicated by the colored triangle,
   provides a first clue towards the importance of each cluster.
   For example, for the $X(3872)$ with $I(J^{PC})=0(1^{++})$ the dominant meson-meson cluster is $DD^\ast$,
   because $DD$ can produce $J=1$ only with non-vanishing orbital angular momentum.\footnote{Note also that
   the parity argument against a pion exchange between $D$ mesons does not apply here because even if the four-body equation
    were simplified to a two-body system like in Fig.~\ref{fig:4body-2body}, the interaction would proceed through quark exchanges.}
   The proximity of the $X(3872)$ mass to the $DD^\ast$ threshold is a strong indication for a dominant $DD^\ast$ component inside this state,
   as has been predicted even before its discovery~\cite{Tornqvist:1993ng,Guo:2017jvc}.
   For the hadrocharmonium component, a $J/\psi$ with $0(1^{--})$
   would require a quantum-number exotic light pseudoscalar $0(0^{--})$ which is not observed.
   This leaves the vector state $\omega$ with $0(1^{--})$ as a possible partner; the threshold $m_{J/\psi}+m_\omega \approx 3880$ MeV
   is again in the same mass region.
   Finally, the sum of the $cq$ scalar ($0^+$) and axialvector ($1^+$) diquark masses
   in the rainbow-ladder truncation is $m_S+m_A = 5.1(1)$ GeV (cf.~Table~\ref{tab:rl}),
   which is more than 1~GeV higher compared to the other thresholds and  suggests again that diquarks are less important.

   In Refs.~\cite{Wallbott:2019dng,Wallbott:2020jzh}
   the four-body BSE was solved in the rainbow-ladder truncation, with the BSA
   simplified to the leading tensors corresponding to the physically relevant two-body clusters
   in each of the three configurations. Upon absorbing the
   kinematical dependence of the BSA stemming from the higher momentum multiplets
   in Eq.~\eqref{momentum-multiplets} into explicit two-body pole factors,
   which depend on the calculated two-body meson and diquark masses,
   only the dependence on the symmetric variable $\mS_0$ remains dynamical.
   This significantly reduces the numerical complexity but at the same time allows
   one to judge the relative importance of the individual two-body clusters.

\begin{figure*}[t]
\includegraphics[width=0.5\textwidth]{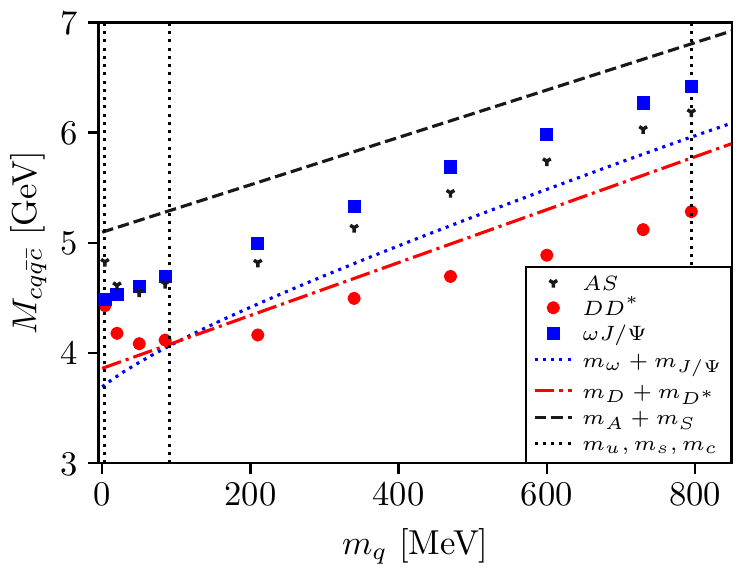}
\includegraphics[width=0.5\textwidth]{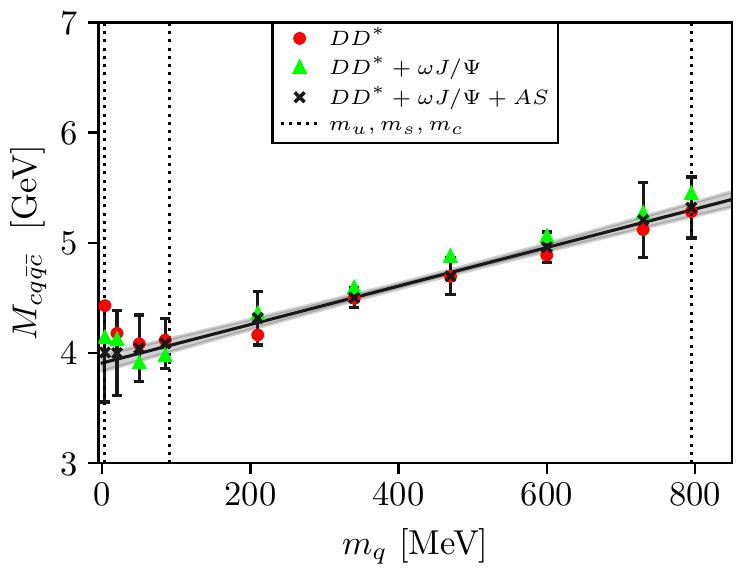}
\caption{Mass of the $I(J^{PC})=0(1^{++})$ four-quark state corresponding to the $X(3872)$ as a function
of the light current-quark mass~\cite{Wallbott:2019dng}. The solutions for the individual
         $DD^\ast$, $J/\psi\,\omega$ and diquark-antidiquark components together with their respective thresholds are shown in the left panel
         and those including one, two and all three channels in the right. The error bars combine the extrapolation error
         with that obtained by varying the momentum partitioning.
         The line with error band is a fit to the data points below threshold.
          }  \label{fig:X}
\end{figure*}

   Fig.~\ref{fig:X} shows the resulting $cq\overline{qc}$ mass in the $0(1^{++})$ channel,
   where the mass of the $c\bar{c}$ pair is fixed and the light current-quark mass $m_q$
   is varied from the charm to the strange and light-quark masses (indicated by vertical dashed lines).
   The left panel displays the results using the dominant tensors
   for each of the three configurations separately --- $DD^\ast$ (red circles), $J/\psi\, \omega$ (blue squares) and $AS$ (black triangles down) ---,
   together with their respective thresholds. The $DD^\ast$ component indeed
   has the lowest mass, followed by the diquark-antidiquark and $J/\psi \,\omega$ components.
   Note that the dynamical $J/\psi \,\omega$ result lies far above its threshold even though the thresholds
   for $J/\psi \,\omega$ and $DD^\ast$ are similar.
   The tetraquark mass obtained with the diquark-antidiquark component lies below its threshold,
   whereas the one obtained with $DD^\ast$ is below threshold only
   above the strange-quark mass. The right panel shows the results for $DD^\ast$ only (red), $DD^\ast + J/\psi \,\omega$ (green)
   and finally all three components (black). One can see that the addition of these
  tensors has little effect and the mass of the state is
  essentially determined by the $DD^\ast$ component alone.

  On a technical note, the masses in Fig.~\ref{fig:X} are obtained by calculating the eigenvalues $\lambda_i(P^2)$ of the kernel of the four-quark BSE
  in Fig.~\ref{fig:4b-bse} at each quark mass $m_q$ and below the lowest-lying two-body threshold in each setup, i.e., for $P^2 > - M_\text{thr}^2$ along the real axis.
  For a bound state, the condition $\lambda_i(P^2)=1$ is met on the real axis below the threshold, whereas for a resonance it holds
  in the complex plane of $P^2$ on a higher Riemann sheet.
  In principle one could extract the resonance pole locations by solving the four-quark BSE for complex momenta on the first sheet
  using contour deformations in the integration, in combination with analytic continuation  methods~\cite{Tripolt:2016cya,Tripolt:2018xeo,Williams:2018adr,Binosi:2019ecz,Eichmann:2019dts,Santowsky:2020pwd},
  although this comes with a substantial numerical effort.
  So far, unless the resulting mass is below all relevant thresholds and can be inferred directly, the eigenvalue curve $\lambda_i(P^2)$ is extrapolated from the real axis
  using rational functions to obtain an estimate for the real part of the resonance mass.
  In addition, the error stemming from the reduction of the kinematical variables discussed above can be estimated
   by varying the momentum partitioning between the quark momenta which add up to the total momentum $P$.
  The combined error from the extrapolation and varying the momentum partitioning leads to the error bars in the right panel of Fig.~\ref{fig:X}.
  Finally, a linear fit is performed using the results at higher quark masses below threshold;
  this results in the error band shown in the plot.
  The resulting mass at the physical $u/d$ point is $M_{1^{++}}^{cq\overline{qc}} = 3916(74)$ MeV,
  in good agreement with the mass of the $X(3872)$.

\begin{figure*}[t]
\includegraphics[width=0.5\textwidth]{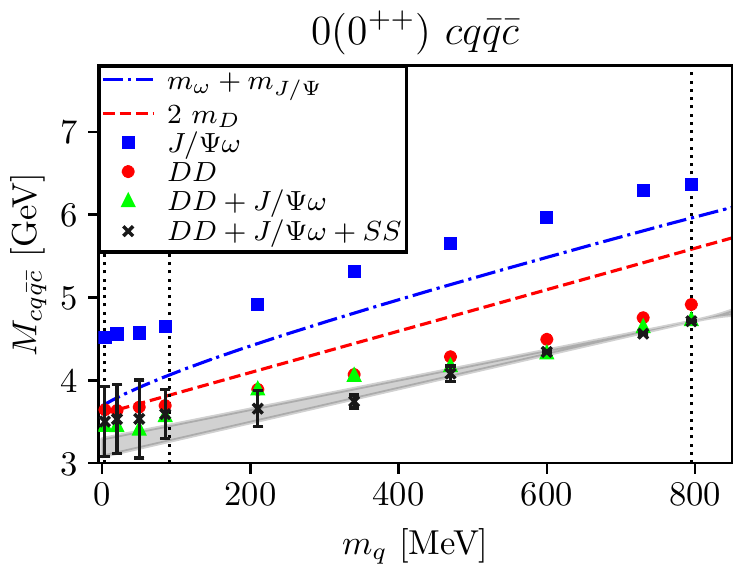}
\includegraphics[width=0.5\textwidth]{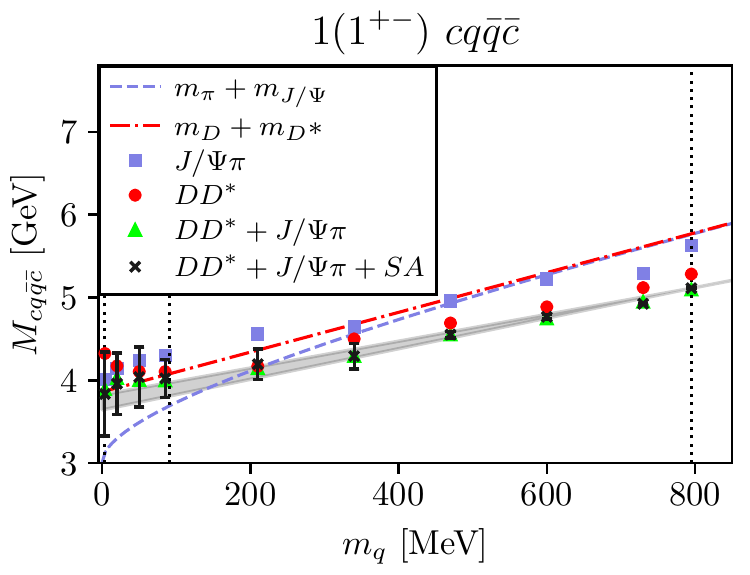}
\caption{Quark-mass evolution of the $cq\overline{qc}$ ground states in the $0(0^{++})$ and $1(1^{+-})$ channels
             for different components of the four-body amplitude (same notation as in Fig.~\ref{fig:X}).~\cite{Wallbott:2020jzh}
          }  \label{fig:XZ}
\end{figure*}

  The analogous analysis for the $I(J^{PC})=0(0^{++})$ and $1(1^{+-})$ channels with quark content $cq\overline{qc}$ is shown in Fig.~\ref{fig:XZ}.
  The former is (likely) carried by the $X(3915)$ and the latter by the $Z_c(3900)$.
  Also here the pattern that arises is very similar in both cases: the meson-meson components ($DD$ in the scalar and $DD^\ast$ in the axialvector case)
  are the dominant ones, whereas the hadrocharmonium and diquark-antidiquark contributions are subleading.
  The effect is however less pronounced compared to the $X(3872)$.

  Similar observations also hold for $cs\overline{sc}$ states: In the $0(1^{++})$ channel the calculations
  produce a state at $M=4.07(6)$ GeV, which is again dominated by meson-meson contributions ($D_s D_s^\ast$)
  and could be identified with the $\chi_{c1}(4140)$, even though it is not particularly close to the $D_s D_s^\ast$ threshold.
  A corresponding state is also found
  in the $0^{++}$ channel, although the large theoretical errors make a prediction of its mass rather imprecise~\cite{Wallbott:2020jzh}.

  In the case of open-charm states with quark content $cc\overline{qq}$,
  the two-body clusters are rather different since charge conjugation is replaced by Pauli antisymmetry, see Eqs.~(\ref{pauli-12}--\ref{pauli-34}).
  Instead of a hadrocharmonium component one has two identical heavy-light meson components
  plus a configuration with a heavy diquark and a light antidiquark,
   where the heavy scalar $cc$ diquark is forbidden by Pauli antisymmetry:
   \begin{itemize}\setlength\itemsep{1mm}
   \item (13)(24) and (14)(23), $(c\bar{q})(c\bar{q})$, molecule: $(D,D^\ast,\dots)\times(D,D^\ast,\dots)$
   \item (12)(34), $(cc)(\overline{qq})$, diquark-antidiquark: $A \times (A,S)$
   \end{itemize}
   In this case the diquark contributions are no longer negligible but significant~\cite{Wallbott:2020jzh}.
   The resulting four-quark masses in the $0(1^+)$ and $1(1^+)$ channels lie in the 4 GeV region and are in the ballpark obtained with other approaches~\cite{Eichten:2017ffp,Luo:2017eub,Karliner:2017qjm}.

   Finally, in analogy to our strategy pursued for the light-quark sector
   it will be interesting to compare the results for heavy-light four-quark states with those obtained in the two-body approach;
   a corresponding study is already underway.

\section{Summary and outlook}
\label{sec:5}

 In this article we have summarized existing results within the functional approach
 to four-quark states using Dyson-Schwinger and Bethe-Salpeter equations.
 We discussed the mechanism how a dynamical generation of resonances emerges from the quark level in
 terms of internal two-body clusters, and we capitalized on the potential of the approach to distinguish
 different two-body clusters in terms of molecular-like, hadrocharmonium and diquark-antidiquark components.
 Certainly, work on this subject has only begun and there are many open issues that need to be addressed in future studies:
 the calculation of widths, decay modes and matrix elements;
 the inclusion of more complete tensor bases and momentum dependencies;
 the mixing with $q\bar{q}$ components; the study of four-quark states with bottom quarks;
 and the implementation of three- and four-body interactions.
 Over the last decades, exotic meson spectroscopy has become a cutting-edge area of research
 in experimental hadron physics,
 and similar progress in theory will be essential to understand and interpret existing and forthcoming data.

\begin{acknowledgements}
We are grateful to Evgeny Epelbaum, Christoph Hanhart, Soeren Lange, Sasa Prelovsek, George Rupp, Marc Wagner and Richard Williams for discussions.
This work was supported by the DFG grant FI 970/11-1 and the FCT Investigator Grant IF/00898/2015.
\end{acknowledgements}

\bibliographystyle{spphys}
\bibliography{X_tetraquark3}

\end{document}